\documentclass[a4paper,11pt]{article}
\usepackage{pos_ILD}

\usepackage[T1]{fontenc}
\usepackage{siunitx}
\usepackage{mathtools}
\DeclarePairedDelimiter\abs{\lvert}{\rvert}%
\usepackage{hyperref}
\usepackage[capitalise]{cleveref}
\usepackage{graphicx}
\usepackage{subcaption}
\usepackage{array} 
\usepackage{booktabs}
\usepackage[section]{placeins}

\usepackage{lineno} 

\usepackage[british]{babel} 
\usepackage{csquotes} 

\title{Development of time-of-flight particle identification for future Higgs factories}
\ShortTitle{TOF PID for future Higgs factories}

\author*[a,b]{Bohdan Dudar}
\author[a]{Ulrich Einhaus}
\author[a]{Jenny List}
\author[a,c]{Konrad Helms}
\author[a]{Frank Gaede}

\affiliation[a]{Deutsches Elektronen-Synchrotron DESY, Notkestr. 85, 22607 Hamburg, Germany}

\affiliation[b]{Universität Hamburg, Hamburg, Germany}

\affiliation[c]{Georg-August-Universität Göttingen, Göttingen, Germany}

\emailAdd{bohdan.dudar@desy.de}
\emailAdd{ulrich.einhaus@desy.de}
\emailAdd{jenny.list@desy.de}
\emailAdd{konrad.helms@desy.de}
\emailAdd{frank.gaede@desy.de}

\abstract{
With the emergence of advanced Silicon (Si) sensor technologies such as LGADs, it is now possible to achieve exceptional time measurement precision below 50 ps. As a result, the implementation of time-of-flight (TOF) particle identification (PID) for charged hadrons at future $e^{+}e^{-}$ Higgs factory detectors has gained increasing attention. Other PID techniques require a gaseous tracker with excellent dE/dx resolution, or a Ring-imaging Cherenkov detector (RICH), which adds additional material in front of the calorimeter.
TOF measurements can be implemented either in the outer layers of the tracker or in the electromagnetic calorimeter, and are thus particularly interesting as a PID method for detector concepts based on all-silicon trackers and optimised for particle-flow reconstruction.
In this study, we will explore potential integration scenarios of a TOF measurement in a future Higgs factory detector, using the International Large Detector (ILD) as an example. We will focus on the challenges associated with crucial components of TOF PID, namely track length reconstruction and TOF measurements. The subsequent discussion will highlight the vital impact of precise track length reconstruction and various TOF measurement techniques, including recently developed machine learning approaches. We will evaluate the performance in terms of $\pi/K$ and $K/p$ separation as a function of momentum, and discuss potential physics applications.}

\FullConference{The European Physical Society Conference on High Energy Physics (EPS-HEP2023)\\
 21-25 August 2023\\
Hamburg, Germany\\}

\begin{document}
\maketitle

\section{Introduction}
The particle physics community is in an active discussion about the next collider facility, an $e^{+}e^{-}$ collider for precision studies of the Higgs boson, the top quark and the electroweak gauge bosons, as recommended by the current European strategy for particle physics from 2020 and the recent Snowmass community study.
Alongside the effort to converge on an optimal collider proposal, there is ongoing development of different detector designs for such future Higgs factories.
One open question for such detectors is the implementation of fast $\mathcal{O}(\qty{10}{\ps})$ timing into the detector design.
Fast timing is interesting for several aspects: time-of-flight (TOF) particle identification (PID), 4D tracking, 5D calorimetry, background rejection and long-lived particle searches.
This study focuses on the TOF PID of $\pi$, $K$ and $p$ using the International Large Detector (ILD)~\cite{ILD} as an example case.
We present the latest developments of the reconstruction algorithms necessary for TOF PID, discuss involved challenges and present their performance.
Additionally, we briefly touch on implications for detector concepts with different main tracker technologies and potential physics applications of the TOF PID.

\section{Basics of time-of-flight particle identification}

For a simple helix track, the mass $m$ of a particle can be expressed in terms of its momentum $p$, its TOF $T$ and its track length $L$, as shown in \cref{eq:mass}:

\begin{equation}
    m = p\sqrt{\frac{c^{2}T^{2}}{L^{2}} - 1} \text{.}
    \label{eq:mass}
\end{equation}

In a realistic detector, energy loss and scattering lead to distortions from a simple helix, and the relation becomes more complicated~\cite{winnilength}.
However, \cref{eq:mass} remains a good approximation if appropriate definitions of $p$ and, more importantly, $L$ are used.
Instead of the momentum of the particle at some specific point, we use the weighted harmonic mean, as defined in \cref{eq:hm_momentum}:

\begin{equation}
    \small p = \sqrt{\langle p^{2} \rangle_{HM}} =\sqrt{ \sum^{n}_{i=1} L_{i} \bigg/ \sum^{n}_{i=1} \frac{L_{i}}{p^{2}_{i}}} \text{.}
    \label{eq:hm_momentum}
\end{equation}

The sum runs over all $n$ track segments, with $L_{i}$ being $i\text{th}$ segment length and $p_{i}$ being the momentum of the particle reconstructed at the beginning of the given $i\text{th}$ track segment.
In practice, using the harmonic mean instead of the momentum of the particle at the IP has a negligible impact.

\section{Track length reconstruction}

Historically TOF PID was usually limited by the TOF resolution.
However, at future Higgs factories with $\mathcal{O}(\qty{10}{\ps})$ time resolutions, the track length reconstruction can become a leading uncertainty factor which limits the PID capabilities.
In our study, the reconstruction of the track length from IP to the ECAL is based on the track parametrization chosen in ILD, which is based on five parameters: curvature $\Omega$, helix dip angle $\tan{\lambda}$, initial azimuthal angle $\varphi_{0}$ and impact parameters $d_{0}$ and $z_{0}$ given with respect to a reference point~\cite{track_parameters}.
For a long time, the ILD reconstruction calculated the track length using \cref{eq:old_len}:

\begin{equation}
    L = \bigg | \frac{ \varphi_{ \text{ECAL}} - \varphi_{\text{IP} } }{ \Omega_{\text{IP}}} \bigg | \sqrt{1 + \tan^{2}{\lambda_{\text{IP}}}}\text{.}
    \label{eq:old_len}
\end{equation}

In this study, we introduce an improved method to reconstruct the track length, defined in \cref{eq:new_len}.

\begin{equation}
    L = \sum_{i=1}^{n} L_{i} = \sum_{i=1}^{n} \frac{ |z_{i+1} - z_{i}| }{| \tan{\lambda}_{i} |} \sqrt{1 +\tan^{2}{\lambda}_{i}}
    \label{eq:new_len}
\end{equation}

\begin{figure}[!htb]
\centering
\begin{subfigure}[t]{.35\textwidth}
  \centering
  \includegraphics[width=1.\linewidth]{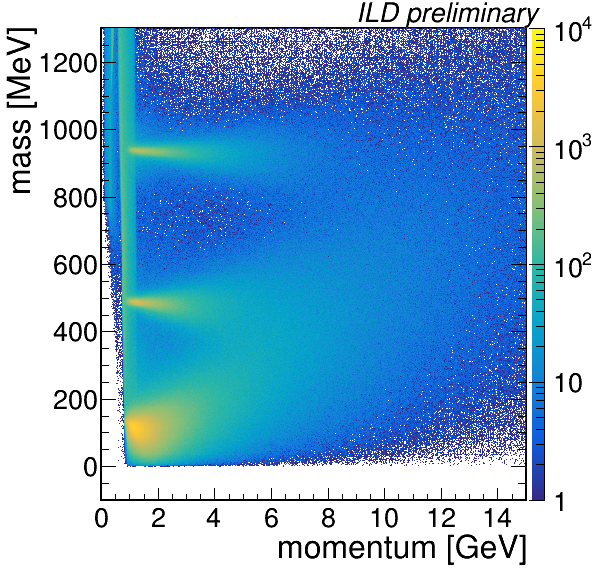}
  \caption{Old track length (\cref{eq:old_len}) used in~\cite{ILD}.}
  \label{fig:old_len}
\end{subfigure}\hspace{2cm}
\begin{subfigure}[t]{.35\textwidth}
  \centering
  \includegraphics[width=1.\linewidth]{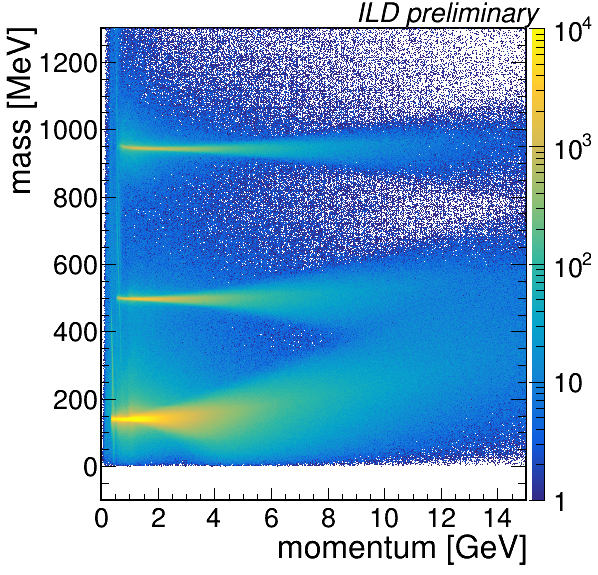}
  \caption{New track length (\cref{eq:new_len}).}
  \label{fig:new_len}
\end{subfigure}
\caption{The mass of charged hadrons reconstructed from TOF as a function of their momentum for previous and improved track length reconstruction. In both cases, perfect time information is assumed for illustration.}
\label{fig:trk_len}
\end{figure}

Both \cref{eq:old_len,eq:new_len} are equivalent for any perfectly helical track.
However, on the fully reconstructed particles from the full ILD simulation Monte-Carlo (MC) samples, one can see a significant improvement of the mass reconstruction in \cref{fig:trk_len}. This is due to the fact that fully reconstructed tracks are not exactly perfect helices and experience numerous effects such as energy loss as well as track fitting uncertainties.

\Cref{fig:trk_len} shows that a new approach using \cref{eq:new_len} results in significantly better-reconstructed masses compared to the previous approach using \cref{eq:old_len}.
The improvement comes from three major factors.
Firstly, \cref{eq:old_len} reconstructs a helix length based on only two points, which does not work for tracks with multiple curls, as expected for low momentum particles.
Secondly, it has been observed that using $\tan{\lambda}$ and $z$ instead of $\Omega$ and $\varphi$ as base parameters of the track evolution results in better performance since the former tends to be reconstructed with better precision than the latter.
Lastly, \cref{eq:new_len} uses multiple hits along the track summing over the lengths of many small helix pieces, which are built from the local helix track parameters at every tracker hit utilizing the full information of the Kalman filter, that is used in the track fit.
This accounts for energy loss effects along the track.

\section{Time-of-flight reconstruction}

A dedicated timing layer provides robust TOF measurement, but its effectiveness is constrained by the single measurement time resolution.
Integration of time measurements into the electromagnetic calorimeter (ECAL) potentially improves TOF resolution as $1/\sqrt{N}$, with $N$ being the number of ECAL hits with time information.

In this section, we study TOF reconstruction using the ILD ECAL, a highly granular Si/W sandwich calorimeter with thirty layers of $5 \times 5\, \si{\mm^{2}}$ cells~\cite{ILD}.
We consider three scenarios for the different time measurement implementation scenarios in ILD assuming different hit time resolutions:
1) using only the first ECAL layer with a \qty{30}{\ps} time resolution;
2) utilising the double strip layer of ILD Silicon External Tracker (SET) with a \qty{50}{\ps} hit time resolution;
3) using hits in the first ten ECAL layers, selecting in each layer only the closest hit to the track's extrapolation line in the ECAL with \qty{100}{\ps} hit time resolution.

\begin{figure}[!htb]
\centering
\begin{subfigure}[t]{.35\textwidth}
  \centering
  \includegraphics[width=1.\linewidth]{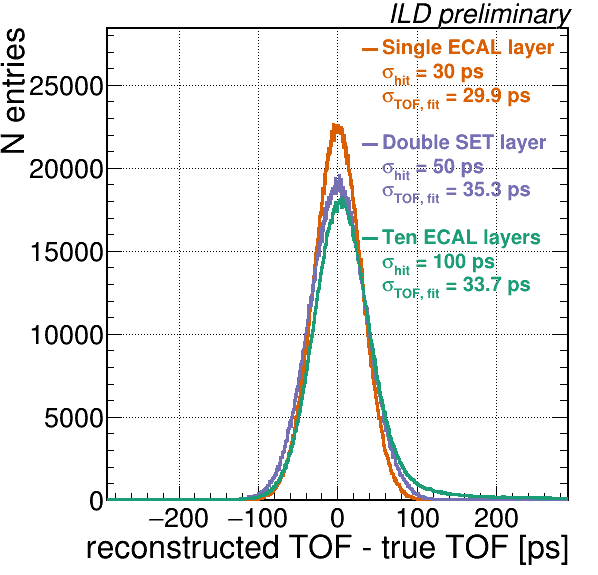}
  \caption{Linear scale.}
  \label{fig:tof_res_lin}
\end{subfigure}\hspace{2cm}
\begin{subfigure}[t]{.35\textwidth}
  \centering
  \includegraphics[width=1.\linewidth]{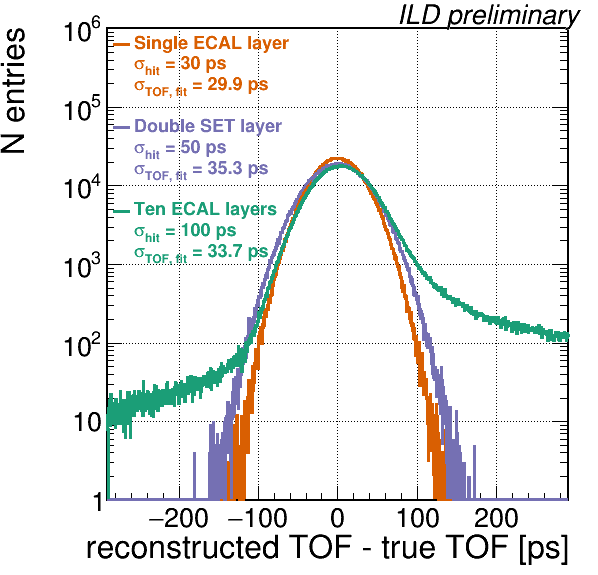}
  \caption{Log scale.}
  \label{fig:tof_res_log}
\end{subfigure}
\caption{TOF resolution obtained using three different hypothetical scenarios of time measurement integration in ILD using a single ECAL layer, double SET layer or ten ECAL layers with different hit time resolutions.}
\label{fig:tof_res}
\end{figure}

In all cases, we correct the ECAL hits' time information by their distance from the track’s extrapolated entry point into the ECAL, divided by the speed of light.
The reconstructed TOF is the average of corrected time measurements.
We compare this to the MC true TOF information, which is defined as the corrected MC truth time of the closest hit to the track’s extrapolated entry point into the ECAL.
The performance of these methods is presented in \cref{fig:tof_res}.

The third method, highlighted in green, shows that it is possible to use multiple hits from in-depth of the ECAL shower to get $\approx \qty{33.7}{\ps}$ TOF resolution using \qty{100}{\ps} hit time resolution if one does a hit preselection to ensure that hits further away from the track don't introduce a significant bias.
Despite this preselection, the third method still has distinctive non-Gaussian tails, visible in \cref{fig:tof_res}.
Using a machine learning (ML) approach for the TOF reconstruction could provide significant improvements over the conventional method by using the full information to its utmost potential.
We develop an ML model based on the EPiC encoder~\cite{epic_regression} followed by a multi-layer perceptron using all spatial, time and energy information of the ECAL hits as well as reconstructed track parameters, such as momentum, length and extrapolated entry point into the ECAL.
Though the ML model is still under development, the first results look already promising.

\section{Separation power for time-of-flight particle identification}
The PID performance has been studied for various TOF resolutions with the latest track length reconstruction.
The separation power between particle hypotheses is evaluated following the method of~\cite{sep_power}:
The overlap between the reconstructed mass distributions for pions and kaons is calculated.
Then two normal distributions with $\mu_1$, $\mu_2$ and $\sigma := \sigma_1 = \sigma_2$ are placed at a distance $\abs{\mu_1 - \mu_2}$ with the same overlap value, leading to a well-defined separation power $S = \abs{\mu_1 - \mu_2} / \sigma$.

\begin{figure}[!htb]
\centering
\begin{subfigure}[t]{.35\textwidth}
  \centering
  \includegraphics[width=1.\linewidth]{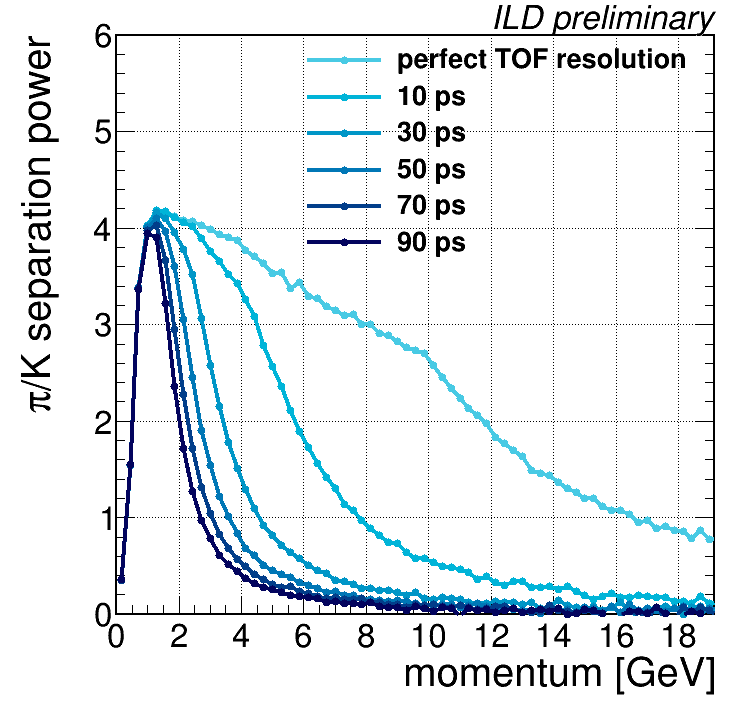}
  \caption{$\pi$/$K$ separation.}
  \label{fig:sep_pik}
\end{subfigure}\hspace{2cm}
\begin{subfigure}[t]{.35\textwidth}
  \centering
  \includegraphics[width=1.\linewidth]{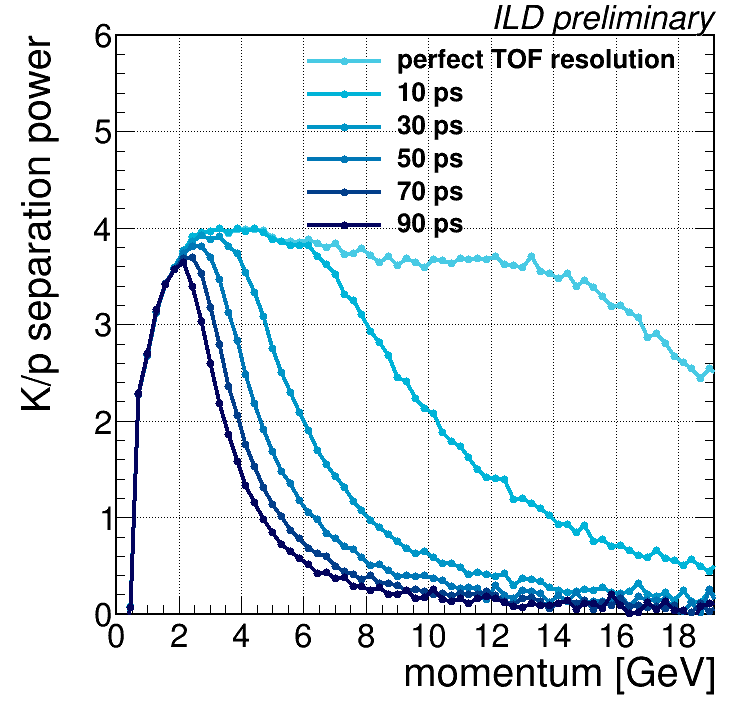}
  \caption{$K$/$p$ separation.}
  \label{fig:sep_kp}
\end{subfigure}
\caption{The separation power between charged hadrons particles based on TOF PID, with the latest track length reconstruction and different TOF resolutions as a function of the momentum.}
\label{fig:sep}
\end{figure}

\Cref{fig:sep} shows that with perfect TOF resolution, only limited by the track length reconstruction, one could achieve $\pi$/$K$ ($K$/$p$) separation at the 3$\sigma$ level for momenta up to \qty{8}{\GeV} (\qty{17}{\GeV}). This shows a significant potential if the TOF resolution per particle could be improved beyond \qty{10}{\ps}, e.g. by ML means utilising multiple independent time measurements from the ECAL, as discussed in the previous section. In the case of a single time measurement with a time resolution of \qty{50}{\ps} the momentum reach of TOF PID becomes significantly limited reaching $\pi$/$K$ ($K$/$p$) separation at the 3$\sigma$ level up to \qty{2.1}{\GeV} (\qty{3.7}{\GeV}).

\section{Potential physics applications}

Generally, kaon-ID is a crucial tool for the precision measurements planned at a future collider facility.
TOF PID works on a limited low-momentum range and nicely complements other PID tools such as dE/dx, which perform best at high momentum and have a notable 'blind spot' at low momentum.
The most common processes at $E_{\text{CM}} = \qty{250}{\GeV}$ $Z \rightarrow q\bar{q}$ and $WW \rightarrow q\bar{q}q\bar{q}$ contain a sizable fraction of charged hadrons at low momentum, which is covered by TOF easily. For illustration, about 38\% (51\%) of the $K$ ($p$) have a momentum below \qty{3}{\GeV}, which should be identifiable with TOF assuming \qty{30}{\ps} TOF resolution, which is deemed feasible with state-of-the-art technology with a single time measurement.
Enhanced PID capabilities of a detector improved by TOF PID have the potential to contribute to various precision studies in flavour physics, Higgs sector, $q/\bar{q}$ separation for $A_{\mathrm{FB}}$, $V_{\mathrm{cs}}$ and $V_{\mathrm{cb}}$ measurements, as well as to enhance event reconstruction, extend long-lived particle searches and measure the charged kaon mass.
A quantitative assessment of TOF PID contributions for various physics studies is a subject of further studies.

\section{Summary}

We introduced a novel track length reconstruction method, highlighting its importance for mass reconstruction.
Our study presents the results for the ML-based TOF estimator, using time information in the first ten ECAL layers with a \qty{50}{\ps} hit resolution. This technique yields a TOF resolution at the distribution core $\approx \qty{6.7}{\ps}$, outperforming the previous state-of-the-art analytical algorithm.
We studied the impact of TOF resolution on the momentum reach for particle separation using our novel track length estimator. A TOF resolution per particle better than \qty{30}{\ps} significantly improves the momentum reach, particularly as it approaches \qty{0}{\ps}. A resolution worse than \qty{30}{\ps} provides limited momentum reach, diminishing as the resolution worsens.
TOF could nicely complement a PID system of a detector at a future Higgs factory and potentially enhance precision studies reliant on PID. However, a quantitative assessment of the benefits requires future studies.

\section{Acknowledgements}

We would like to thank the LCC generator working group and the ILD software working group for providing the simulation and reconstruction tools and producing the Monte Carlo samples used in this study.
This work has benefited from computing services provided by the ILC Virtual Organisation, supported by the national resource providers of the EGI Federation and the Open Science GRID. In this study we widely used the National Analysis Facility (NAF)~\cite{Haupt_2010}
and would like to thank Grid computational resources operated at Deutsches Elektronen-Synchrotron (DESY), Hamburg, Germany.
We thankfully acknowledge the support by the Deutsche Forschungsgemeinschaft (DFG, German Research Foundation) under Germany's Excellence Strategy EXC 2121 "Quantum Universe" 390833306.

\bibliographystyle{JHEP}
\bibliography{refs}
\end{document}